\documentclass[
aps,%
12pt,%
final,%
notitlepage,%
oneside,%
onecolumn,%
nobibnotes,%
nofootinbib,%
superscriptaddress,%
showpacs,%
centertags]%
{revtex4-1}

\usepackage[english]{babel}
\usepackage{graphicx}

\newcommand{\JP}{\psi}
\newcommand{\beq}{\begin{equation}} \newcommand{\eeq}{\end{equation}}
\newcommand{\beqa}{\begin{equation*}} \newcommand{\eeqa}{\end{equation*}}
\newcommand{\hS}{\hat s} \newcommand{\hT}{\hat t} \newcommand{\hU}{\hat u}
\newcommand{\hsig}{\hat\sigma}
\newcommand{\eps}{\epsilon}

\begin{document}

\selectlanguage{english}

\author{A.V. Luchinsky}
\email{Alexey.Luchinsky@ihep.ru}
\affiliation{Institute for High Energy Physics, Protvino, Russia}
\author{S.V. Poslavsky}
\email{stvlpos@mail.ru}
\affiliation{Institute for High Energy Physics, Protvino, Russia}

\title{Inclusive charmonium production at PANDA experiment}

\begin{abstract}
The production of the charmonium states in $p \bar p$ experiments is considered
at the energy rates near threshold in NLO. Such a consideration allows one to
obtain 
non zero distributions over the transverse momentum of the final charmonium and
gives a natural explanation to the existence of $\chi_{c1}$-meson in final
state, that is observed experimentally and cannot be produced in leading order
processes. The crucial role of scale parameter choice in $\alpha_s (Q^2)$ and
partonic distributions $f(x,Q^2)$ shown, and the correct choice offered.  
\end{abstract}

\maketitle

\section{Introduction}
Precision experimental studies of the charmonium production in proton-antiproton
collisions at low energies proposed in new experiment, labeled PANDA
\cite{PANDA}. Such studies allows to obtain a deeper understanding of charmonium
physics. PANDA experiment deals with  proton target and antiproton beam energies
up to $15$GeV. This energies lies near charmonium production threshold. In
present article we give a theoretical predictions of the inclusive charmonium
production in $p\bar p$ collisions by accounting next to leading order diagrams
in partonic cross sections. Such an approach was considered by a number of
authors \cite{70GEV,Baier:1981uk,Gastmans:1987be,Meijer:2007} and applied mainly
for proton-proton collisions.

 The important part of the Program is the detailed analysis of all possible
mechanisms of charmonia production. Such an analysis is especially important
since at low energies there is significant difference between charmonium
production in  $p \bar p$ and $p p$. For the $p p$ at low energies the
contributions of gluon-gluon, quark-gluon and quark-antiquark subprocesses are
comparable, while for $p \bar p$ the quark-antiquark annihilation subprocess
dominates. For example, if the energy of the proton beam is equal to 40 GeV, the
ratio of $\JP$ production cross sections in $p\bar p$ and $pp$ collisions equals
$\sigma(p\bar p)/\sigma(pp)\sim6$. 
 
Another problem is that the direct production of $\JP$-meson is suppressed in
comparison with the production of the intermediate $P$-wave states
$\chi_{c0,1,2}$ with the subsequent decay $\chi_c\to J/\psi\gamma$. This fact is
well confirmed in the experiments \cite{Alexopoulos:1999wp}. The experimentally
observed cross sections of $\chi_{c2}$ and $\chi_{c1}$ are comparable
($\chi_{c0}$-meson can hardly be observed due to its small radiative width),
while the well known Landau-Yang theorem forbids the formation of the axial
meson from two massless gluons. One more difficulty is that the partonic
distributions are integrated over the transverse momentum. As a result, such
method does not allow one to obtain the distributions of $\chi_{c0}$- and
$\chi_{c2}$-mesons over $p_T$.

Initially this problems were solved by the introduction of color-octet (CO)
components of the quarkonia, that arise naturally in the non-relativistic
quantum chromodynamics (NRQCD), where the expansion over the relative velocity
of quarks in meson is performed. In this model it is assumed, that final meson
is formed from heavy quark pair in color-octet state, that subsequently
transforms into a physically observed colorless meson. In the framework of NRQCD
the probabilities of these transitions are described by the matrix elements of
four-fermion operators, that are determined from the experimental distributions
over the transverse momentum of final charmonium.  We would like to stress,
however, that this explanation will not work for charmonium production at lower
energies. The reason is that the distributions caused by octet components
decreases slowly with the rise of the transverse momentum, but the probability
to find such a component in the meson is small, compared with the singlet case.
As a result, in the large transverse momentum region the contribution of octet
components can be significant, but for small energies and transverse momenta it
is suppressed.

Recently another way to solve this problem was proposed, where the so called
non-integrated over the transverse momentum distribution functions $G(x,k_T)$
are used ($k_T$-factorization) \cite{Teryaev:1996sr,Kniehl:2006sk,Baranov:1}. In
this case both mentioned above problems are solved simultaneously . The
transverse momentum of the produced in gluon fusion $\chi_{c0,2}$ mesons is
explained by the transverse momenta of the initial partons. Axial charmonium
meson can also be produced in gluon fusion, since in the framework of $k_T$
factorizations gluons have non-zero virtuality of the order $k_T^2$. There is a
number of works, that explain the experimental distributions on TEVATRON with
the help of these functions (see, for example
\cite{Hagler:2000dd,Likhoded:2006xu,Kniehl:2006sk}). According to these works,
there is no need to introduce CO components to reproduce the experimental data
on $P$-wave charmonium $p_T$ distributions. Thus, in $k_T$-factorization
approach color-singlet components give the main contribution.

Unfortunately, the method, used in the modeling of the unintegrated distribution
functions $G(x,k_T)$, is based on the summation of large $\log(1/x)$, so it is
not applicable for low energies, where the gluon momentum fractions are in the
range $0.1<x_g<0.5$. For this reason we are forced to use the following
approximation in our calculations. We start from the collinear gluon
distributions with well known collinear distribution functions. Further we
consider the charmonia production at next to leading (NLO) order in the strong
coupling constant $\alpha_s$. Such a trick enables us to obtain the
distributions over $p_T$ for all charmonium states. For $\chi_{c0}$ and
$\chi_{c2}$ production we observe a collinear singularity at $p_T=0$. To avoid
this singularity we introduce a regularization procedure. For directly produced
$\JP$ and $\chi_{c1}$ such a singularity is absent and we use the whole
integration region for $p_T$.

Next section is devoted to the consideration of different modes of charmonia
production and collinear singularities regularization. In the third section we
determine the cross sections of the hadronic processes and pay attention to the
correct scale parameters choice in $\alpha_s(Q^2)$ and parton distributions
$f(x,Q^2)$. There we are present numerical results. 

\section{Partonic subprocesses}

\begin{figure}
\includegraphics[scale=0.7]{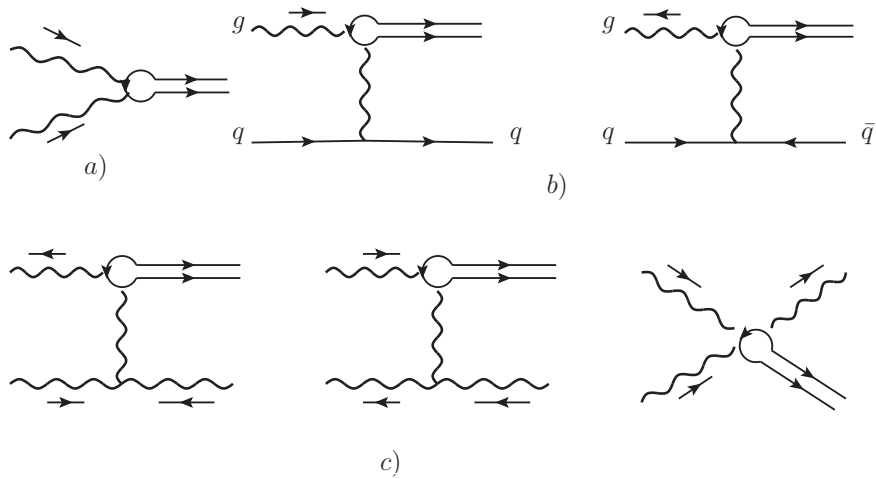}
\caption{ Feynman diagrams for the charmonium production. a) direct
$\chi_{c0,2}$ production. b) through $qg\to\mathcal{Q}q$ and  $\bar q
q\to\mathcal{Q}g$ subprocesses, where $\mathcal{Q} = \chi_{cJ}$. c) through
$gg\to\mathcal{Q}g$ subprocess, where $\mathcal{Q} = \chi_{cJ}$ for the first
two diagrams and $\mathcal{Q} = \JP,\chi_{cJ}$ for the last diagram.}
\label{diagrams}
\end{figure}

Feynman diagrams corresponding to the charmonium production are presented in
Fig.\ref{diagrams}. At the leading (Fig.\ref{diagrams}a) order only processes
$gg \to \chi_{c0,2}$ are available. Process $gg \to \JP$ is forbidden by charge
parity. Process $gg\to \chi_{c1}$ is forbidden due to Landau-Yang theorem, that
forbids the formation of the axial meson from two massless gluons. Other LO
cross sections we give in Appendix \ref{reg_cross}.

Next to leading order diagrams are presented in Fig.\ref{diagrams}b,c. First two
diagrams in Fig.\ref{diagrams}c includes 3-gluon vertex. To avoid delicate
problems, bounded with ghosts contributions, we recalculated differential cross
sections for process $gg \to Qg$ ($Q = \JP, \chi_{c0,1,2}$) (and also for $qg\to
\chi_{0,1,2} q$) in axial gauge:
\begin{equation}
\label{L_gf}
{\mathcal L}_{gf} = - \frac{1}{2\xi} (n_{\mu} A^{a \, \mu})^2, \mbox{   with } 
n_{\mu} n^{\mu} = -1  
\end{equation}
%
%
This gauge does not require additional ghosts Lagrangian, but gluon propagator
and polarization sum becomes more complicated. Using $\xi = 0$ (Landau choice),
it can be found that
\begin{eqnarray}
\sum \eps_{\mu}(k) \eps_{\nu}(k) &=& -\eta_{\mu\nu} +
\frac{k_{\mu}k_{\nu}}{(k,n)^2} - \frac{k_{\mu}n_{\nu}+k_{\nu}n_{\mu}}{(k,n)},
\nonumber\\
D_{\mu\nu}^{ab} (k) &=& \delta^{ab} \frac{1}{k^2} \sum \eps_{\mu}(k)
\eps_{\nu}(k)
\label{polarization_sum}
\end{eqnarray}
Auxiliary vector $n_{\mu}$ will disappear in physical observables. We found,
that our results are in excellent agreement with \cite{Gastmans:1987be},
\cite{Meijer:2007} for $gg$ channel and  \cite{70GEV} for $qg$ channel. Exact
formulas for the differential cross sections are rather tedious and can be found
in the cited papers. 

Massless $\hT$-channel gluon in propagators in Fig.\ref{diagrams} leads to the
collinear singularities in small $\hat t$  and $\hat u = M^2- \hS -\hT$ regions,
so  $gg \to \chi_{0,2} g$ and $qg \to \chi_{0,2} q$ are divergent:
\begin{equation}
\frac{d \hat \sigma}{d  \hat t}  \sim \frac{1}{\hat t \hat u}.
\end{equation}
In terms of meson transverse momentum $p_T = \sqrt{\hT\hU/\hS}$, these
singularities correspond to $p_T \to 0$ singularity. To calculate the total
cross section: 
\begin{equation}
 \hsig (\hS) = \int\limits_{M^2-\hS}^{\hT} d \hT \,\, \frac{d \hsig(\hS,\hT)}{ d
\hT} =  \int\limits_{0}^{(\hS -M^2)/2\sqrt{\hS}} d p_T \,\, \frac{d
\hsig(\hS,p_T)}{ d p_T},
\end{equation}
some regularization should be performed. Popular decision is to cut off the
small $p_T$ region, i.e. restrict integration region in last formula by setting
$p_T > \Delta$, where cutoff parameter $\Delta$ can be  taken from experimental
setup or from some physical reasons. For example in \cite{70GEV}, $\Delta$ was
taken equal to $1/R_{c \bar c}$, where $R_{c \bar c}$ is the geometrical size of
charmonium. Such approach have a big lack, because the total cross section is
high sensitive to $\Delta$ variation. To avoid these difficulties, we will use
another regularization procedure.

Taking indefinite integral of differential cross section, the following well
known relation can be found:
\beq
\label{p_ggg}
\int d \hT \,\, \frac{d \hsig (gg \to Q g)}{ d \hT}  = \frac{\alpha_s}{2
\pi}\hsig_0(gg \to Q) P_{g\to gg} \left(\frac{M^2}{\hS}\right)
\ln{\frac{\hU}{\hT}} + \mbox{finite}
\eeq
Similarly, for $qg\to Q q$ reaction we have
\beq
\label{p_qgq}
\int d \hT \,\, \frac{d \hsig (qg \to Q q)}{ d \hT}  = \frac{\alpha_s}{2
\pi}\hsig_0(gg \to Q) P_{q\to qg} \left(\frac{M^2}{\hS}\right)
\ln{\frac{\hU}{\hT}} + \mbox{finite}.
\eeq
In these expressions the  second parts are finite at $\hT \to 0$ and $\hT \to
M^2-\hS$, while $\hsig_0(gg \to Q)$ are given in (\ref{LO_CHI0}),
(\ref{LO_CHI1}). $P_{g\to gg} (x) $ and $P_{q\to qg} (x)$ are well known QCD
splitting functions:
\begin{eqnarray*}
P_{g\to gg} (x) &=& 6\left( \frac{x}{1-x} + \frac{1-x}{x} + x(1-x) \right)\\
P_{q\to qg} (x) &=& \frac{4}{3} \frac{1+(1-x)^2}{x}
\end{eqnarray*}
The full hadronic cross section can be obtained by integrating partonic cross
sections with partonic distribution functions:
\beq
\label{total_sig}
\sigma(s) = \int d x_1 d x_2 f(x_1) f(x_2) \hsig(\hS)
\eeq
Singular parts of (\ref{p_ggg}) and (\ref{p_qgq}) are included in partonic
distributions functions $f(x)$ and generate well known scaling violations,
described by Altarelli-Parisi equations. So, inclusion of singular parts in
partonic cross sections leads to double-counting: one time in $\hsig$ and one in
$f(x)$. Thus, we will use regularization:
\beq
\label{reg_gg}
\hsig^{Reg}(gg \to Q g)  = \left.\left(\int \frac{d \hsig (gg \to Q g)}{ d \hT}
\,d \hT  - \frac{\alpha_s}{2 \pi}\hsig_0(gg \to Q) P_{g\to gg}
\left(\frac{M^2}{\hS}\right) \ln{\frac{\hU}{\hT}} \right)
\right|_{\hT=M^2-\hS}^{\hT=0}
\eeq
and similarly:
\beq
\label{reg_qg}
\hsig^{Reg}(qg \to Q q)  = \left.\left(\int \frac{d \hsig (qg \to Q q)}{ d \hT}
\,d \hT  - \frac{\alpha_s}{2 \pi}\hsig_0(gg \to Q) P_{q\to qg}
\left(\frac{M^2}{\hS}\right) \ln{\frac{\hU}{\hT}} \right)
\right|_{\hT=M^2-\hS}^{\hT=0}
\eeq

 In contrast to $\chi_{c0,2}$, differential cross sections for $\JP$ and
$\chi_{c1}$ have no collinear singularities and finite at $\hT \to 0$ and $\hU
\to 0$. In case of $\chi_1$, this is explained by the Landau-Yang  theorem, that
forbids the production from two massless gluons. As a result, the squared matrix
element of this reaction is proportional to the virtuality of the intermediate
t-channel gluon, so this factor compensates the divergency, caused by the
propagator. For $\JP$, we have similar reasoning based on charge parity, since
the first two diagrams of Fig.\ref{diagrams}c are absent in this case.

Exact formulas for regularized partonic cross sections are given in Appendix
\ref{reg_cross}

\section{Hadronic cross sections}

Let us now consider full hadronic process
\beq
 A(P_1)  B(P_2) \to \mathcal{Q} (P) + X,
\eeq
where A and B are the initial hadrons, $\mathcal{Q} = \JP, \chi_{cJ}$ , and in
the parentheses corresponding particle momenta introduced. The cross section of
this reaction is expressed through the cross sections
of considered above partonic reactions :
\beq
\sigma(s) = \sum_{a,b} \int d x_1 d x_2 f_{a/A}(x_1) f_{a/B}(x_2)
\hsig_{ab}(\hS),
\eeq
where summation is performed over partons $a$ and $b$, $x_{1,2}$ are the
momentum fractions held by these partons, and $f_{a/A} (x_1 )$, $f_{b/B} (x_2)$
are the distribution functions of the partons in the initial hadrons. In the
common variables
\begin{eqnarray}
x &=& x_1-x_2\\
\label{define_x}
\hS &=& (x_1 P_1 +x_2 P_2)^2 = x_1 x_2 s,
\end{eqnarray}
the full hadronic cross-section becomes
\beq
\label{sig_total}
\sigma(s) = \sum_{a,b} \int\limits_{M^2}^s d \hS \,\, \hsig_{ab}(\hS)
\int\limits_{-x(\hS)}^{x(\hS)} \left. \frac{d x}{\tilde x} f_{a/A}(x_1)
f_{b/B}(x_2) \right|_{x_{1,2} = x_{1,2}(x,\hS)},\,\,\,\,\,\tilde x=
x_1+x_2,
\eeq
where
\beq
\label{limit_x}
x (\hS) = 1- \frac{\hS}{s}
\eeq
In our numerical estimates we used the distribution functions and $\alpha_s$
numeric values presented in the work \cite{Alekhin:2002fv}. Other numerical
parameters are equal to:
\begin{eqnarray}
&M_{\JP} = 3.097 \,\mbox{GeV},\,\, &M_{\chi_{c0}} =  3.415\,\mbox{GeV}, \\
&M_{\chi_{c1}} =  3.511\,\mbox{GeV},\,\, &M_{\chi_{c2}} =  3.556\,\mbox{GeV}\\
&R_{S}^2 (0) = 0.81 \,\mbox{GeV}^3,\,\,  &{R'}_{P}^2 (0) = 0.075 \,\mbox{GeV}^5
\end{eqnarray}

\subsection{Scale dependence}

\begin{figure}
\includegraphics[scale=0.8,height=1.8in]{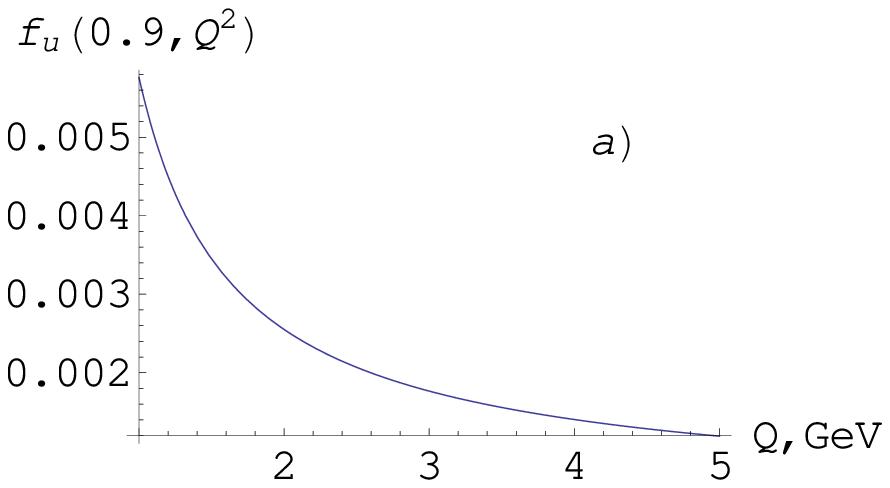}
\includegraphics[scale=0.8,height=1.8in]{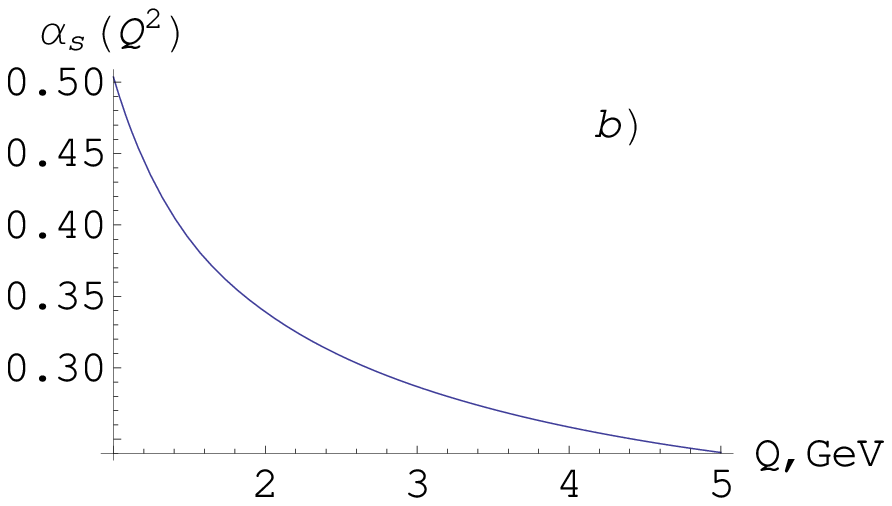}
\caption{ a)  $Q$-dependence of $u$-quark distribution at $x$ = 0.9 b) 
$Q$-dependence of $\alpha_s$}
\label{scale}
\end{figure}

There are two physical quantities, that depend on some scale choice: partons
distributions $f_{a/A}(x, Q^2)$ and strong coupling constant $\alpha_s(Q^2)$.
The parton distribution function $f_{a/A}(x, Q^2)$, gives the probability of
finding a parton $a$ of longitudinal fraction $x$ in physical (anti)proton,
taking into account collinear gluon (or massless quark) with transverse momenta
$p_T < Q$. It is clear, that exact value of $Q$ depends on parameters of
partonic subprocess, i.e. $Q^2 = Q^2(\hS)$. It is convenient to set $Q^2$ to a
fixed value $Q^2_*$ - characteristic momentum transfer of partonic subprocess.
Such choice can be argued by mean value theorem, which states: 
\begin{eqnarray*}
\int\limits_{a}^{b}  f(x) g(x) d x = f(x_*) \int\limits_{a}^{b}  g(x) dx 
\end{eqnarray*}
On the other hand, from the structure of Altarelli-Parisi equations, it is
clear, that at least for high energies ($ Q \gg 1$ GeV) error in choosing of
$Q_*$ leads to negligibly small variation of final results; for example, the
parton distributions change by $\sim 1 \%$  as $Q_*$ is changed by a factor of
10. Another situation we have at low energies ($Q \ll 1$ GeV), when perturbation
theory works bad, and error in choosing of $Q_*$ leads to a dramatic variation
of partonic functions. At the energy rates $\sim 1$ GeV,  error in $Q_*$ in two
times can leads to $f(x, Q^2_*)$ error about 20\%. For example, u-quark
distribution dependence on $Q_*$ is shown on Fig.\ref{scale} a).

Second physical quantity, depending on some scale choice is $\alpha_s(Q^2)$.
Here, the meaning of $Q$ is different from the meaning in distribution
functions. The strong coupling $\alpha_s$ scale dependence occurs when the full
propagators and vertexes are inserted in tree level diagrams. Similar reasoning,
based on mean value theorem allows to set  $\alpha_s(Q^2)$ to a fixed value with
some scale $Q_*$. The perfect justification of this procedure is given in
\cite{Brodsky:1983}. Only in simplest situations the exact value of $Q_*$ can be
found: for example for $2\to2$ reaction through $s$-channel particle, it can be
found, using Callan-Symanzik renorm-group equation, that $Q_*^2 = s$. In other
cases the exact value of scale can be found only by sequential analysis of
perturbation series expansion. It is clear, that $Q_*$ depends on the process.
At high energies $Q \gg 1$ GeV $\alpha_s(Q^2)$ becomes almost constant. At the
energies rates near 1 GeV, $\alpha_s$ dependence of $Q$ is shown on
Fig.\ref{scale} b).

So, in general, we have three possible ways of setting scale parameters in full
cross section:
\begin{eqnarray*}
\label{fixed_scheme}
\mbox{fixed scheme} \,\,\,\,\,\,\sigma(s,Q^2_*) &= \sum_{a,b}
\int\limits_{M^2}^s d \hS \,\, \hsig_{ab}(\hS,\alpha_s(Q^2_*))
\int\limits_{-x(\hS)}^{x(\hS)} \frac{d x}{\tilde x} f_{a/A}(x_1,Q^2_*)
f_{b/B}(x_2,Q^2_*) \\
\label{float_scheme}
\mbox{float scheme}\,\,\,\,\,\,\sigma(s,Q^2_*) &= \sum_{a,b} \int\limits_{M^2}^s
d \hS \,\, \hsig_{ab}(\hS,\alpha_s(\hS)) \int\limits_{-x(\hS)}^{x(\hS)} \frac{d
x}{\tilde x} f_{a/A}(x_1,Q^2_*) f_{b/B}(x_2,Q^2_*) \\
\label{float2_scheme}
\mbox{float2 scheme}\,\,\,\,\,\,\sigma(s,Q^2_*) &= \sum_{a,b}
\int\limits_{M^2}^s d \hS \,\, \hsig_{ab}(\hS,\alpha_s(\hS))
\int\limits_{-x(\hS)}^{x(\hS)}  \frac{d x}{\tilde x} f_{a/A}(x_1,\hS)
f_{b/B}(x_2,\hS) 
\end{eqnarray*}
The fixed scheme is the most common way, used in calculations. The float scheme,
takes in account the fact, that $\alpha_s$ in partonic subprocess, depends on
$\hS$. Of course, in general, this dependence is complicated and has the form
$\alpha_s(f(\hS))$, but from general considerations it is clear, that for small
interval, near process threshold, $f(\hS) \sim \hS$. The float2 scheme, takes in
account both $\alpha_s(Q^2)$ and $f(x,Q^2)$ scaling. It is also clear that
maximum transverse momentum $Q$ in $f(x,Q^2)$ depends on the $\hS$ and for small
energy intervals we set $Q^2 \sim \hS$. Actually, the cross section in float2
scheme does not depends on $Q_*$.

\begin{figure}
\includegraphics[scale=0.75]{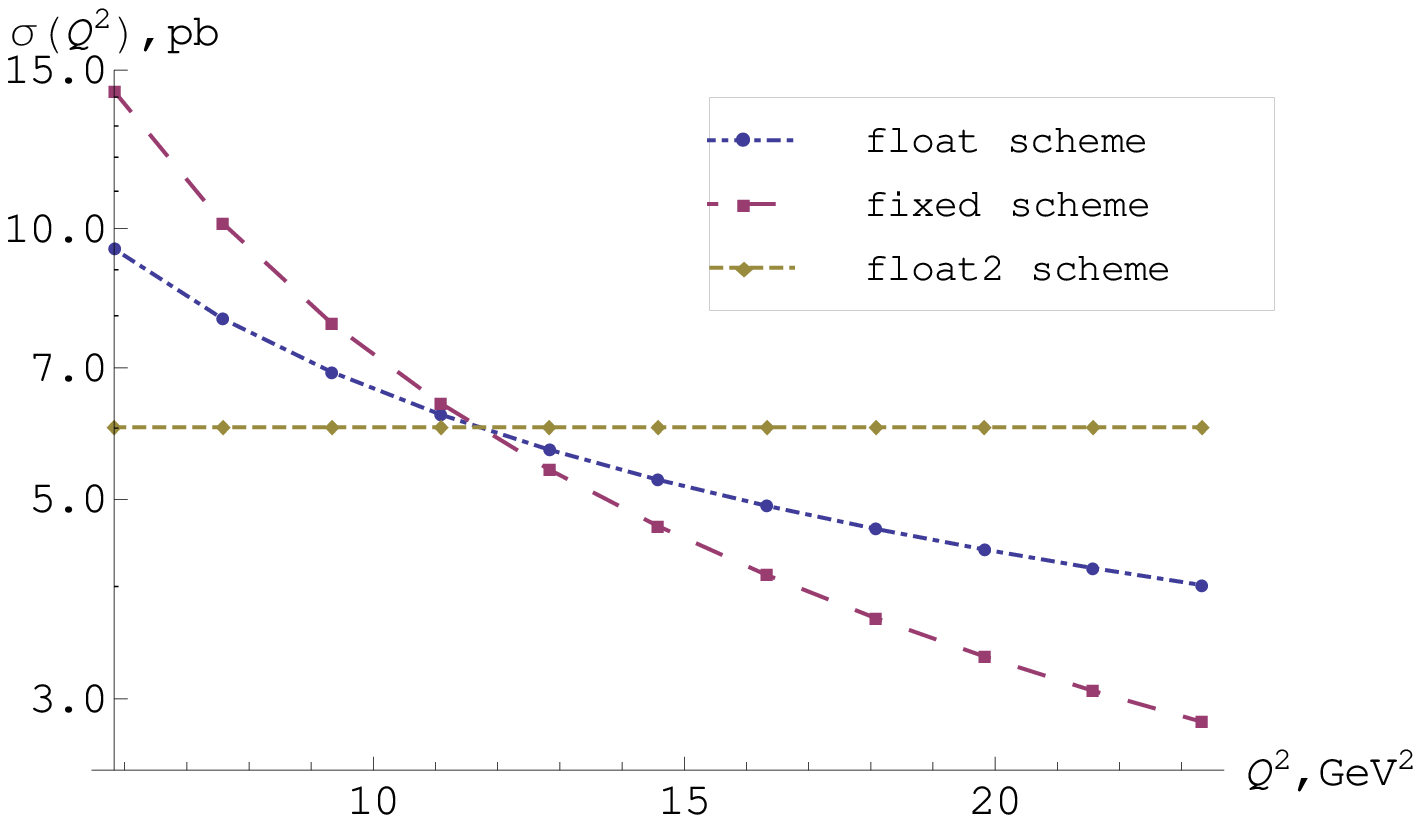}
\caption{Scale dependence in different schemes of the $u \bar u$ channel cross
section in $p \bar p \to \chi_{c0} X$ collision at the energy $\sqrt{s} = 4.34
\mbox{GeV}.$}
\label{schemes}
\includegraphics[scale=0.75]{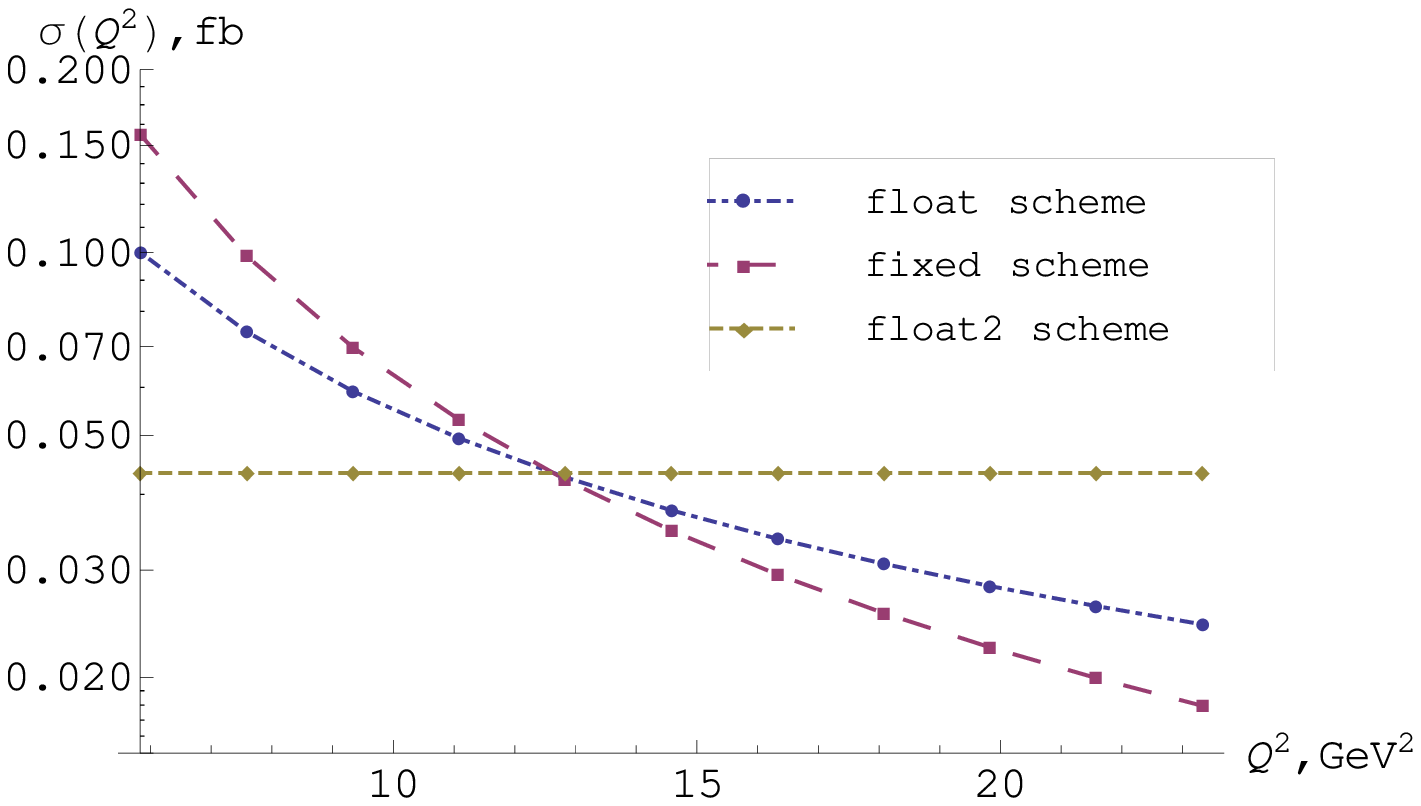}
\caption{Scale dependence in different schemes of the $u g$ channel cross
section in $p \bar p \to \chi_{c0} X$ collision at the energy $\sqrt{s} = 4.34
\mbox{GeV}.$}
\label{schemesb}
\end{figure}

Fig.\ref{schemes} illustrates the dependence of the cross section on scheme
choice for $\chi_{c0}$ production in $u\bar u$ channel at $p \bar p$ collisions.
It is seen, that all three curves are crossed in one point at $Q^2 =
M^2_{\chi_{c0}} = 11.66\mbox{GeV}^2$.  For other mesons and other channels the
picture is equivalent - three curves are crossed at the corresponding squared
meson mass. The only exception --- quark-gluon channel, where cross point
greater at 10\%, then corresponding squared meson mass (Fig.\ref{schemesb}). We
make similar calculations for other energy regions and found, that this results
remain valid. As was excepted, at high energies ($s \gg 1 \mbox{GeV}^2 $) the
difference between schemes is negligible. So, this results improves that all
schemes are equivalent with the correct $Q_*$ choice:
\beq
Q_* = M,
\eeq
where $M$ - corresponding meson mass. 

In all further calculations we shall use the fixed scheme with $Q_* = M$.

\subsection{Total cross sections}

\begin{figure}
\includegraphics[scale=0.83]{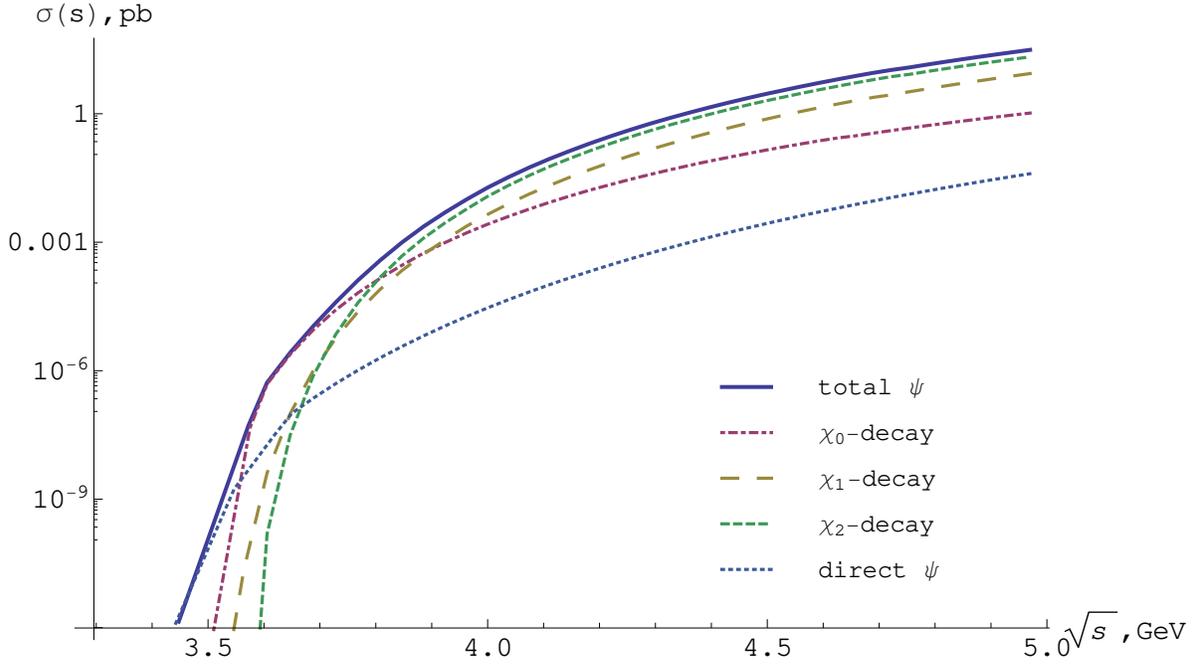}
\caption{Different contributions to the total $\psi$ production in
proton-antiproton reaction for different c.m. energies. 1 --- total $\JP$
production, 2 --- $\JP$ production through the radiative $\chi_{c2}$ decay, 3
--- through the radiative $\chi_{c1}$ decay, 4 --- through the radiative
$\chi_{c0}$ decay, 5 --- direct $\JP$.}
\label{final}
\end{figure}

Fig.\ref{final} illustrates the dependence of $\JP$ production through different
processes on total energy. The bold line shows the summed over all processes
cross section:
\begin{equation}
\label{eq:sigPsi}
   \sigma =
\sigma_{gg}(\JP)+\mbox{Br}(\chi_{c0}\to\JP\gamma)\sigma(\chi_{c0})+\mbox{Br}
(\chi_{c1}\to\JP\gamma)\sigma(\chi_{c1})+\mbox{Br}(\chi_{c2}
\to\JP\gamma)\sigma(\chi_{c2}), 
\end{equation}
where the branching values are equal to:
\begin{equation}
\mbox{Br}(\chi_{c0}\to\JP\gamma) = 0.016,\,\,\,\mbox{Br}(\chi_{c1}\to\JP\gamma)
= 0.344,\,\,\,\mbox{Br}(\chi_{c2}\to\JP\gamma) = 0.195
\end{equation}
 From this picture it is seen, that direct $\JP$ production and production from
radiative $\chi_{c0}$ decay is highly suppressed. The contribution of
$\chi_{c0}$ radiative decay is negligible, due to very small branching value. As
was noted above, the direct $\JP$ production only available in process $gg \to
\JP g$, but the gluon-gluon channel in $p \bar p$ reactions is highly suppressed
by the quark-antiquark channel, so the cross section of the direct $\JP$
production is significantly smaller, then the production of $\chi_{cJ}$, where
the quark-antiquark channel is available.

\begin{figure}
\includegraphics[scale=0.78]{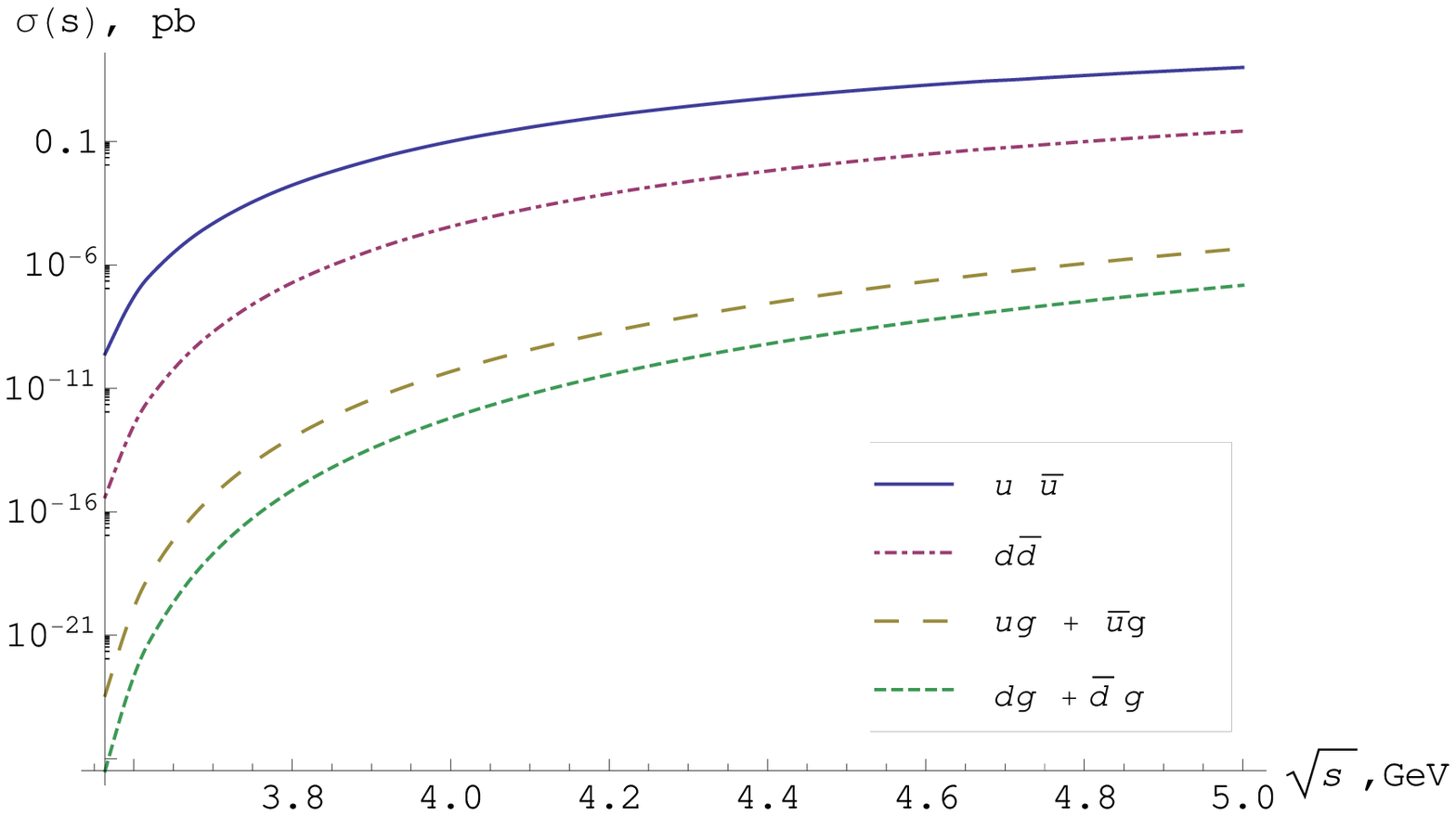}
\caption{Contribution of the different subprocesses to the total $\chi_{c1}$
production. 1 --- $u \bar u$ subprocess, 2 --- $d \bar d$ subprocess, 3 --- the
sum over $u g$ and $\bar u g$ subprocesses, 4 --- the sum over $d g$ and $\bar d
g$ subprocesses.}
\label{sub1}
\includegraphics[scale=0.73]{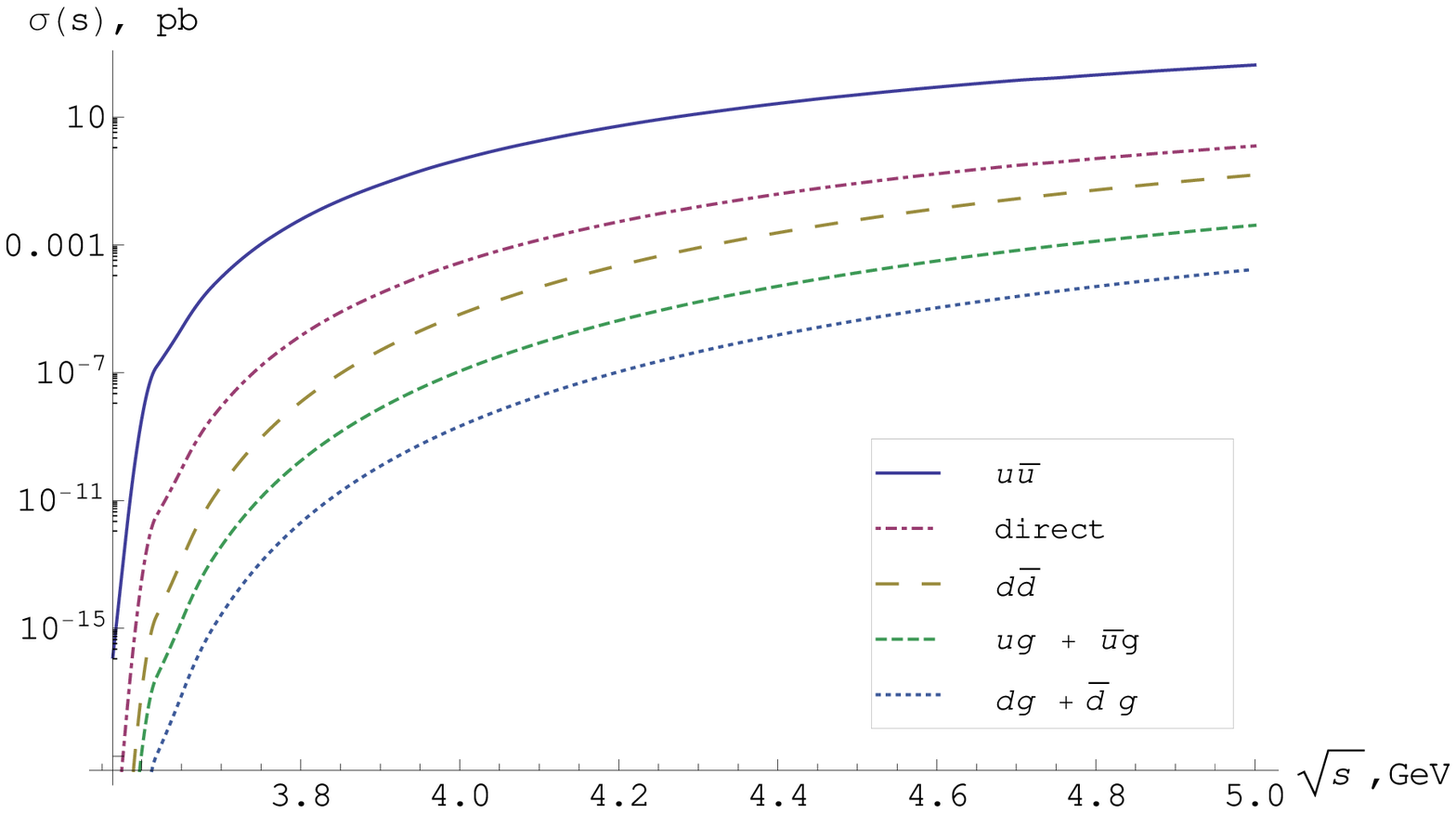}
\caption{Contribution of the different subprocesses to the total $\chi_{c2}$
production. 1 --- $u \bar u$ subprocess, 2 --- direct $\chi_{c2}$, 3 --- $d \bar
d$ subprocess, 4 --- the sum over $u g$ and $\bar u g$ subprocesses, 5 --- the
sum over $d g$ and $\bar d g$ subprocesses.}
\label{sub2}
\end{figure}

In the Fig.\ref{sub1} and Fig.\ref{sub2} we show the contributions of the
different subprocesses to the total $\chi_{c1}$ and $\chi_{c2}$ production cross
sections. For both mesons the most significant contribution gives the $u \bar u$
subprocess. For the gluon-gluon subprocess our numerical results are equals to
zero within the error of numerical calculations. The negligibly small effect of
the other channels can be easily explained by the structure of the parton
distributions. The small energy  of the hadronic reaction corresponds to the
longitudinal fraction $x$ in partonic distributions $f_{a/A}(x)$ close to zero.
For this region the $u$-quark distribution function absolutely dominates.

\subsection{Production mode at the $s =  32 \mbox{ GeV}^2$}

It should be stressed, that presented above expressions for charmonia production
cross sections can be considered only as estimates on upper bounds. The reasons
is that in some events initial (anti)protons are present also in the final
state. Due to baryonic number conservation in proton-proton scattering this
configuration is realized almost always. In proton-antiproton interaction,
however, that presence of baryons in the final state is not necessary.
Numerically this effect can be described in terms of inelasticity coefficient,
which can be interpreted as the probability of proton-antiproton annihilation
into other states. According to presented in \cite{Ding:1992sx} analysis, this
coefficient is equal to $K\sim0.5$ and decreases slightly with the increase of
energy.

If initial baryons are present also in the final state, the effective
interaction energy decreases from $\sqrt{s}$ to $\sqrt{s_\mathrm{eff}} \sim
\sqrt{s} - 2M_p$. In the case of high-energy colliders this modification does
not change significantly the cross sections of the considered processes. For
PANDA environment, however, situation is completely different. From
fig.\ref{final} it is clear, that the decrease from $\sqrt{s} \sim 5.5$ Gev to
$\sqrt{s_\mathrm{eff}} \sim 3.5$ GeV leads to dramatic decrease of charmonia
production cross sections. So, one can expect, that the reactions $p\bar p\to
p\bar p+J/\psi+X$ give negligible contributions to the cross sections of
charmonia production at PANDA, and expression (\ref{sig_total}) should be
multiplied by the inelasticity factor $K\sim 0.5$.

At the production mode in PANDA experiment the antiproton beam energy equals to
$15\, \mbox{GeV}$, that corresponds to the $s$ value equals to $
32\,\mbox{GeV}^2$. The cross section of $\JP$ meson production is given in
(\ref{eq:sigPsi}). Our calculations give
\begin{equation}
\sigma (p\bar p \to\JP X) = 0.21\,\mbox{nb},
\end{equation}
where
\begin{eqnarray*}
\sigma(p\bar p \to \chi_{c1} X) = 0.2\,\mbox{nb}\\
\sigma(p\bar p \to \chi_{c2} X) = 0.75\,\mbox{nb}\\
\sigma(p\bar p \to \chi_{c0} X) = 0.35\,\mbox{nb}.
\end{eqnarray*}
The ratio of $\chi_{c1}$ and $\chi_{c2}$ production cross sections is equal to
\beq
\frac{\sigma(\chi_{c1})}{\sigma(\chi_{c2})} = 0.26
\eeq
The $p_T$ distribution of mesons, can be obtained by rewriting differential
cross sections in terms of $p_T = \sqrt{\hT \hU/\hS}$ and integrating with the
partonic distributions:
\begin{equation}
\frac{d \sigma}{d p_T} =  \int\limits_{(p_T+\sqrt{p_T^2+M^2})^2}^s \frac{d
\hS}{s}\frac{d \hsig(ab \to \mathcal{ Q} c)} {d
p_T}\int\limits_{-(1-\frac{\hS}{s})}^{1-\frac{\hS}{s}} \frac{d x}{\tilde x}
f_{a/A}(x_1) f_{b/B}(x_2)
\end{equation}
where
\begin{eqnarray*}
\frac{d \hsig}{d p_T} &=& \frac{2 \hS p_T}{\sqrt{(\hS-M^2)^2-4\hS p_T^2}}
\left(\left.\frac{d \hsig}{d \hT}\right|_{\hT =\hT_1} + \left.\frac{d \hsig}{d
\hT}\right|_{\hT =\hT_2}\right),\\
 \hT_{1,2} &=& \frac{1}{2}(M^2-\hS\pm\sqrt{(\hS-M^2)^2-4\hS p_T^2})
\end{eqnarray*}

\begin{figure}
\includegraphics[scale=0.7]{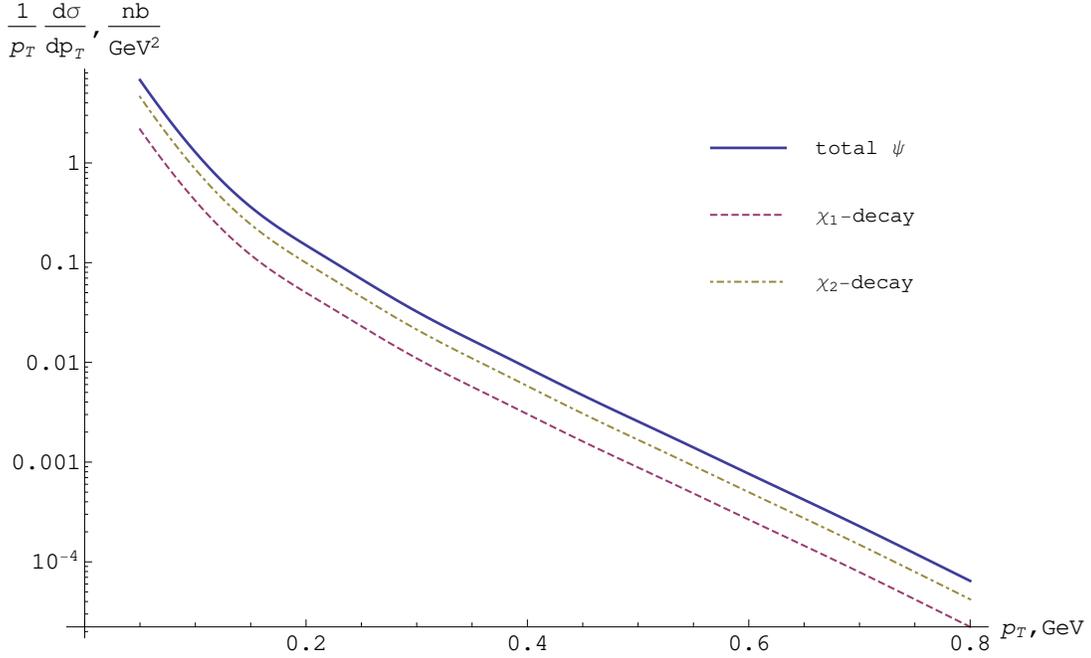}
\caption{Transverse momentum distributions of $\JP$ production with inelesticity
coefficient taken into account. 1 --- total $\JP$, 2 --- $\JP$ production
through $\chi_{c2}$ decay, 3 --- $\JP$ production through $\chi_{c1}$ decay}
\label{pt}
\end{figure}

As was shown before, the major processes, giving contribution to the total $\JP$
production, are the radiative decays of the $\chi_{c1,2}$ mesons, which in turn
are formed through the $u \bar u$ subprocess. Thus, for the calculation of the
$p_T$ dependence we will neglect all other channels of the $\JP$ production. In
thus approximation we do not encounter with collinear singularities, that appear
in other subprocesses of $\chi_{c2}$ formation. Another problem arises when we
consider the total $\JP$ distribution. The radiative decays
$\chi_{cJ}\to\JP\gamma$ can give significant contribution to the
$p_T$-distribution of the final $\JP$, when the transverse momentum $\sim 1$
GeV. However, we will neglect such contribution. On the Fig.\ref{pt} we show the
transverse momentum distributions of the $\JP$ production:
\begin{eqnarray*}
\frac{d \sigma (\chi_{c1,2})}{d p_T} &=& \int \frac{d \hS}{s}\frac{d \hsig(u
\bar u \to  \chi_{c1,2} g)} {d p_T}\int \frac{d x}{\tilde x} f_{u/p}(x_1)
f_{\bar u/\bar p}(x_2),\\
\frac{d \sigma (\JP)}{d p_T} &=& \mbox{Br}(\chi_{c1}\to\JP\gamma) \frac{d \sigma
(\chi_{c1})}{d p_T}+\mbox{Br}(\chi_{c2}\to\JP\gamma) \frac{d \sigma
(\chi_{c2})}{d p_T}
\end{eqnarray*}

\section{Conclusions}

The paper is devoted to $J/\psi$-meson production in proton-antiproton
interaction at low energies. This process can be used to clarify modes of
charmonia production in hadronic experiments and allows one to measure with
higher accuracy proton spectral functions at $x\sim 0.5$.

The physics of charmonia production in hadronic reactions is completely
different for different energies. For high-energy experiments (e.g. Tevatron or
LHC) heavy quarkonia are produced mainly in the gluon-gluon interaction, since
small values of feynman variable $x$ are allowed kinematically. The
contributions of quark-gluon or quark-antiquark modes are negligible. In the
threshold region, where only values $x \sim 0.5$ are allowed, on the contrary,
main contributions come from quark-gluon (in proton-proton interaction) or
quark-antiquark (in the proton-antiproton interactions). In the nearest future
at the FAIR proton-antiproton particle accelerator with $3<\sqrt{s}<5.7$ GeV the
PANDA detector will perform first measurements, so a reliable prediction for
charmonium mesons production for this experiment is required.

In our paper we give predictions for total cross sections of $J/\psi$-meson
production in different modes at NLO. The emission of additional gluon leads to
non-zero transverse momentum of final charmonium, that is obviously absent in LO
partonic reactions $gg\to\chi_{c0,2}$. It is shown, that main contributions to
this process are given by $\chi_{c1,2}$-mesons production due to quark-antiquark
annihilation with the subsequent radiative decay $\chi_{c1,2}\to J/\psi\gamma$. 

Special attention is given to regularization of infrared and collinear
singularities in the case of $\chi_{c2}$ meson production, when the $t$-channel
gluon in $gg\to\chi_{c2}g$ partonic reactions leads to divergency in $p_T$
distribution and infinitive values of the cross section.

\acknowledgments
The authors would like to thank A.K. Likhoded for useful discussions. Also,
authors would like to thank experimentators group: V.V. Mochalov, A.N. Vasiliev
and D.A. Morozov for introduction in PANDA facilities.

This research is partially supported by Russian Foundation for Basic Research
(grant 10-02-00061a). The work of A. V. Luchinsky was also supported by
non-commercial foundation ”Dynasty” and the grant of the president of Russian
Federation for young scientists with PhD degree (grants MK-406.2010.2,
MK-3513.2012.2).

\appendix

\section{Regularized partonic cross sections}
\label{reg_cross}
In this section we give the total cross sections of partonic subprocesses. For
subprocesses, in which the collinear singularities appears, we use the
regularization procedure, described in the main text. 
\subsection{Leading order}

In the leading order only $gg\to\chi_{c0,2}$ reactions are possible. The cross
sections of these reactions are
\begin{equation}
 \hsig (gg\to Q) = \hsig_0 (gg\to Q) \delta(1-M^2/\hS),
\end{equation}
where
\begin{eqnarray*}
\label{LO_CHI0}
\hsig_0 (gg\to \chi_0) = 12 \frac{\pi^2 \alpha^2 {R'}_{\chi}^{2} (0)}{M^5
\hS},\\
\label{LO_CHI1}
\hsig_0 (gg\to \chi_2) = 16 \frac{\pi^2 \alpha^2 {R'}_{\chi}^{2} (0)}{M^5 \hS},
\end{eqnarray*}
where ${R'}_{\chi} (0)$ is  the derivative of the radial part of $\chi$-meson
wave function at the origin,

\subsection{$gg \to Q g$}
As was noted above, the cross sections for processes $gg \to \JP g$ and $gg \to
\chi_{c1} g$ have no collinear singularities, so they does not require
regularization/ We have
\begin{eqnarray*}
\hsig(gg\to\JP g) &=& -\frac{10 \pi  \alpha^3 R_{\JP}(0)^2}{9 \hS^2
\left(M^2-\hS\right)^2 \left(M^2+\hS\right)^3} \left(M^{10}+4 M^8 \hS-2 M^4
\hS^3-M^2 \hS^4-\right.
\end{eqnarray*}
where $R_{\JP}(0)$ is the radial part of $\JP$ wave function at the origin, and
\begin{eqnarray*}
 \hsig(gg\to\chi_1 g) &=& \frac{4 \pi  \alpha^3 {R'}_{\chi}^{2} (0)}{M^7 \hS^2
\left(M^2-\hS\right)^4 \left(M^2+\hS\right)^5} \left(12 M^4 \hS \left(M^{16}+9
M^{14} \hS+\right.\right.
\\ &&
  26 M^{12} \hS^2+\left.28 M^{10} \hS^3+17 M^8 \hS^4+7 M^6 \hS^5-40 M^4 \hS^6-4
M^2 \hS^7-4 \hS^8\right) \log{\frac{M^2}{\hS}}- 
\\ && 
  (M^2-\hS) (M^2+\hS) \left(M^{18}+39 M^{16} \hS+145 M^{14} \hS^2+251 M^{12}
\hS^3+119 M^{10} \hS^4-\right.\\
  && \left.\left.153 M^8 \hS^5-17 M^6 \hS^6-147 M^4 \hS^7-8 M^2 \hS^8+10
\hS^9\right)\right).
\end{eqnarray*}

The $gg$ production of $\chi_{c0,2}$ states, have collinear singularities.
Performing regularization procedure (\ref{reg_gg}), we obtain
\begin{eqnarray*}
\hsig^{Reg}(gg\to\chi_0 g) &=&  
  -\frac{2 \pi  \alpha^3  {R'}_{\chi}^{2} (0) }{3 M^7 \hS^3
\left(M^2-\hS\right)^4 \left(M^2+\hS\right)^5} \,\left(99 M^{24}+132 M^{22}
\hS-7 M^{20} \hS^2-\right.
\\&&
  80 M^{18} \hS^3+210 M^{16} \hS^4-560 M^{14} \hS^5+802 M^{12} \hS^6+696 M^{10}
\hS^7-
\\&&
1721 M^8 \hS^8-244 M^6 \hS^9+
  789 M^4 \hS^{10}+56 M^2 \hS^{11}+12 \hS (-24 M^{22}
\\&&
-41 M^{20} \hS+10 M^{18} \hS^2+7 M^{16} \hS^3+42 M^{14} \hS^4-
  176 M^{12} \hS^5-10 M^{10} \hS^6+
\\&&\left.
40 M^8 \hS^7+14 M^6 \hS^8-31 M^4 \hS^9+9 \hS^{11}) \log{\frac{M^2}{\hS}}-172
\hS^{12}\right),
\\
  \hsig^{Reg}(gg\to\chi_2 g) &=& -\frac{4 \pi  \alpha ^3 {R'}_{\chi}^{2} (0)}{3
M^7 \hS^3 \left(M^2-\hS\right)^4 \left(M^2+\hS\right)^5} \, \left(66 M^{24}+201
M^{22} \hS-\right.
\\&&
  31 M^{20} \hS^2-728 M^{18} \hS^3+360 M^{16} \hS^4-266 M^{14} \hS^5+256 M^{12}
\hS^6+1032 M^{10} \hS^7-
\\&&
752 M^8 \hS^8-271 M^6 \hS^9+207 M^4 \hS^{10}+32 M^2 \hS^{11}-12 \hS \left(12
M^{22}+5 M^{20} \hS+17 M^{18} \hS^2+\right.
\\&&
86 M^{16} \hS^3-204 M^{14} \hS^4+11 M^{12} \hS^5+31 M^{10} \hS^6+74 M^8 \hS^7-8
M^6 \hS^8+22 M^4 \hS^9-
\\&&
\left.\left.6 \hS^{11}\right) \log{\frac{M^2}{\hS}}-106 \hS^{12}\right).
\end{eqnarray*}

\subsection{$qg \to Q q$}
At the $qg$ channel the only $\chi_{cJ}$ mesons can be produced.  The $\chi_1$
meson cross section does not requires regularization and is equals to
\beq
\hsig(qg\to\chi_1 q) = \frac{16 \pi  \alpha ^3 {R'}_{\chi}^{2} (0)}{9 M^7
\hS^3}\left(4 M^6-9 M^2 \hS^2+3 M^4 \hS \log{\frac{\hS}{M^2}}+5 \hS^3\right).
\eeq
Regularized cross sections for $\chi_{c0,2}$ are
\begin{eqnarray*}
\hsig^{Reg}(qg\to\chi_0 q) =& - \frac{16 \pi  \alpha ^3 {R'}_{\chi}^{2} (0)}{27
M^7 \hS^3} \left(4 M^6-18 M^4 \hS+57 M^2 \hS^2+\right.\\
&\left.3\hS \left(4 M^4-9 M^2 \hS+9 \hS^2\right) \log{\frac{M^2}{\hS}}-43
\hS^3\right),\\
\hsig^{Reg}(qg\to\chi_2 q) =& -\frac{16 \pi  \alpha ^3 {R'}_{\chi}^{2} (0)}{27
M^7 \hS^3} \left(20 M^6-36 M^4 \hS+69 M^2 \hS^2+\right.\\
&\left.3 \hS \left(5 M^4-12 M^2 \hS+12 \hS^2\right) \log{\frac{M^2}{\hS}}-53
\hS^3\right).
\end{eqnarray*}

\subsection{$q\bar q \to Q g$}
At the $q \bar q$ channel, all cross sections are finite. This is explained by
the fact, that their differential cross sections are cross-symmetric ($\hT
\leftrightarrow \hS$) to the $qg$ ones:
\beq
   |\mathcal{M}(qg\to Qq)|^2 = \left. |\mathcal{M}(q\bar q\to Qg)|^2
\right|_{\hT \leftrightarrow \hS}
\eeq
and the total cross sections are:
\begin{eqnarray}
\hsig(q\bar q\to\chi_0 g) &=& -\frac{128 \pi  \alpha ^3 {R'}_{\chi}^{2} (0) }{81
M^3\hS^3 \left(M^2-\hS\right)}\left(\hS-3 M^2\right)^2 \\
\hsig(q\bar q\to\chi_2 g) &=& -\frac{256 \pi  \alpha ^3 {R'}_{\chi}^{2} (0) }{81
M^3 \hS^3 \left(M^2-\hS\right)}\left(6 M^4+3 M^2 \hS+\hS^2\right) \\
\hsig(q\bar q\to\chi_1 g) &=& -\frac{256 \pi  \alpha ^3 {R'}_{\chi}^{2} (0)}{27
M^3 s^2 \left(M^2-s\right)} \left(M^2+s\right)
\end{eqnarray}

\end{document}